\documentclass[11pt]{amsart}
\usepackage{amssymb, latexsym, amsmath}
\usepackage{graphicx}
\usepackage{datetime}

\newtheorem{theorem}{Theorem}

\newtheorem{corollary}[theorem]{Corollary}

\newtheorem{question}[theorem]{Question}
 \numberwithin{equation}{section}

\newcommand{\R}{\mathbb R}

\begin{document}
\title[Quasi-local mass]{Four lectures on quasi-local mass}

\author{Mu-Tao Wang}

\begin{abstract} This note is based on a series of four lectures the author gave at University of Montpellier, September 28-30, 2015. He started with 
the notion of mass in general relativity, gave a brief review of some known constructions of quasi-local mass, and then discussed the new quasi-local energy and quasi-local mass which Shing-Tung Yau and the author introduced in 2009. At the end, the proof of the positivity of quasi-local energy was sketched and a stability theorem of critical points of quasi-local energy by Po-Ning Chen, Shing-Tung Yau, and the author was discussed. 

\end{abstract}

\address{Department of Mathematics\\
Columbia University\\ New York\\ NY 10027\\ USA}
\email{mtwang@math.columbia.edu}

\date{\usdate{\today}}

\thanks{Supported in part by  NSF grants DMS-1105483 and DMS-1405152 and by a grant from the Simons Foundation (\#305519 to Mu-Tao Wang). The author would like to thank the organizers, particularly Erwann Delay and Marc Herzlich, of the 
thematic school on geometric aspects of general relativity for their invitation and hospitality. 
}

\maketitle

\section{Introduction}
A fundamental difficulty with mass in general relativity is that there is ``no density for gravitation" by Einstein's equivalence principle. Therefore, the elementary formula
\[\text{mass}=\int_\Omega \text{mass density}\] does not hold in general relativity.  For this and other reasons, most understanding of mass is limited to the total mass of an isolated gravitating system, or an asymptotically flat spacetime, where the mass is evaluated as a flux 2-integral at the asymptotic infinity.  However, a quasi-local description of mass is extremely useful because most physical models are finitely extended regions. On the other hand, once a quasi-local mass definition is available, one can take the limit to define the total mass of an isolated system. In 1982, Penrose \cite{P1} proposed a list of unsolved problems in classical general relativity, and the first was ``find a suitable quasi-local definition of energy-momentum (mass)". 

Here is the plan of my talks:

1. Overview of quasi-local mass in Newtonian gravity and special relativity.

2. What happens in general relativity?

3. The problem of quasi-local mass and expected properties.

4. A brief review of some known quasi-local mass constructions.

5. Introduction to the definition of quasi-local energy and mass of Wang-Yau. 

6. The positivity of quasi-local energy and the stability of an optimal isometric embeddings. 

All geometric objects in this note will be assumed to be smooth unless otherwise is mentioned. 

\section{The notions of mass and energy}
\subsection{Mass in Newtonian gravity and special relativity}

 The fundamental equation in Newtonian gravity is \[\Delta \Phi=4\pi \rho,\] where $\rho$ is the mass density.  For a domain $\Omega\subset \R^3$, the total mass is simply $\int_\Omega \rho$, which can be turned into a boundary integral $\frac{1}{4\pi} \int_{\partial \Omega}\frac{\partial \Phi}{\partial \nu}$. 

 For a matter field in $\R^{3,1}$, the energy-momentum tensor of matter density is a symmetric $(0,2)$ tensor $T_{\mu\nu}$.
 Given a spacelike bounded region (hypersurface) $\Omega$, the energy as seen by an observer $t^\nu$ (future timelike unit) is the flux integral
\[\int_\Omega T_{\mu\nu} t^\mu u^\nu \] where $u^\nu$ is the future timelike unit normal of the hypersurface $\Omega$. The dominant energy condition implies $T_{\mu\nu} t^\mu u^\nu\geq 0$. $T_{\mu\nu}$ also satisfies the conservation law $\nabla^\mu T_{\mu\nu}=0$.

 Suppose $t^\mu$ is a constant future-directed timelike unit vector field (this is a translating Killing field) in $\R^{3,1}$. 
 By the conservation law, $T_{\mu\nu}t^\mu$ is divergence free and thus, \[\int_{\Omega_1} T_{\mu\nu} t^\mu u^\nu =\int_{\Omega_2} T_{\mu\nu} t^\mu u^\nu \] if $\partial\Omega_1=\partial\Omega_2$. 

 Minimizing among all such Killing observers $t^\mu$ gives the quasi-local mass. Moreover, the dual $\int_{\Omega} T_{\mu\nu}u^\nu$ defines the quasi-local energy-momentum 4 vector. Replacing $t^\mu$ by other Killing fields of $\R^{3,1}$ gives the notions of quasi-local angular momentum (rotation Killing field) and quasi-local center of mass (boost Killing field). 

 \subsection{Energy-momentum tensor for gravitation?}
We recall Einstein's field equation   \[Ric_{\mu\nu}-\frac{1}{2} Rg_{\mu\nu}=8\pi T_{\mu\nu}.\] The right hand side $T_{\mu\nu}$ still satisfies the conservation law $\nabla^\mu T_{\mu\nu}=0$ and the dominant energy condition. However, it only accounts for energy contribution from matter fields. For example, the Schwarzschild spacetime is vacuum and thus $T_{\mu\nu}=0$, but there is gravitation mass.

 Einstein's  equation is derived from the variation of the Einstein-Hilbert action on a spacetime domain $D$:
    \[\frac{1}{16\pi}\int_D \,^{(4)}R+\frac{1}{8\pi} \int_{\partial D} K+\int_D L(g, \Phi)\] where $K$ is the trace
    of the second fundamental form of $\partial D$ and $\Phi$ represents the matter fields. One can formally obtain an energy-momentum ${T}_{\mu\nu}^*$ which includes the contribution from gravitation, the so-called Einstein pseudo tensor.  ${T}_{\mu\nu}^*$ is a quadratic expression in terms of first derivatives of the metric $g_{\mu\nu}$ and, as the name suggests, not a tensor in particular. At any point in spacetime, one can choose a normal coordinate system so $T^*_{\mu\nu}$ vanishes at that particular point.  Nevertheless, for an isolated system on which an asymptotically flat coordinate system exists at infinity, the integral of ${T}_{\mu\nu}^*$ gives the total energy, as a flux 2-integral of the surface at infinity. With a right asymptotic assumption, the total energy can be shown to be  
independent of the coordinate system.
 
\subsection{Total mass at infinity}
In general, two types of asymptotic infinity are considered: spatial infinity and null infinity. Spatial infinity is modeled on a time slice $(M, g, p)$ of an asymptotically flat spacetime $N$ where $g$ is the induced metric and $p$ is the second fundamental
form. Outside a compact subset $K$, $M\backslash K$ consists of a union of ends, each of which is diffeomorphic to the complement of a ball in $\mathbb{R}^3$. The diffeomorphism induces an Euclidean
 coordinate system $x^1, x^2, x^3$ on each end. We say $(M, g_{ij}, p_{ij})$ is an asymptotically flat initial data set of order $\alpha>0$ if on each end $g_{ij}-\delta_{ij}=O(r^{-\alpha})$, $p_{ij}=O(r^{-\alpha-1})$ for $r=\sqrt{\sum_{i=1}^3 (x^i)^2}$. Suitable decays on the derivatives of $g$ and $p$ are also assumed. 
For each end, one defines\begin{align} E&=\frac{1}{16\pi} \lim_{r\rightarrow \infty} \int_{S_r} (g_{ij,j}-g_{jj,i}) \nu^i \\
P_i&=\frac{1}{8\pi}\lim_{r\rightarrow\infty} \int_{S_r} \pi_{ij} \nu^j, i=1, 2, 3 \label{adm}\end{align}where $\pi_{ij}=p_{ij}- p^k_{k} g_{ij}$ is the conjugate momentum. In the integrals, $S_r$ denotes a coordinate sphere on an end (with respect to the asymptotically
flat coordinate system) and $\nu^i$ is the outward unit normal.  $(E, P_1, P_2, P_3)$ is the well-known ADM \cite{Arnowitt-Deser-Misner} energy momentum. 

Schoen-Yau \cite{SY1} established the positivity of ADM energy for the first time. The following version is due to the work of Schoen-Yau \cite{SY1, SY2}, Witten \cite{Witten}, Parker-Taubes\cite{PT}, Bartnik \cite{Bartnik1}, Chrusciel\cite{Chrusciel86}, and Eichmair-Huang-Lee-Schoen \cite{EHLS}:

\begin{theorem}
Suppose $(M, g, p)$ is an asymptotically flat initial data set in a spacetime
that satisfies the dominant energy condition. If $\alpha>\frac{1}{2}$
then $E\geq \sqrt{\sum P_i^2}$ and equality holds if and only if $(M, g, p)$ is the data of a spacelike hypersurface in the Minkowski spacetime. 
\end{theorem}
 $m=\sqrt{E^2-\sum P_i^2}$ is the ADM mass. There is also Bondi-Sachs energy-momentum (mass) defined at null infinity. In both cases, the energy is a flux integral on the boundary surface of an end at infinity where gravitation is weak and an asymptotically flat coordinate exists.

\section{The problem of quasilocal mass}

 What happens in the regions where gravitation is strong? It is not possible to choose a reference coordinate system then. All the above suggest that we should look for boundary integrals for the expression of a quasi-local mass.

The problem of quasi-local mass can be stated in the following: in a physical spacetime, for each spacelike 2-surface that bounds a spacelike region, define the notion of quasi-local mass and energy-momentum. Properties that are expected to satisfy are:

(1) Positivity under dominant energy condition for a large class of surfaces. 

(2) The quasi-local mass should be zero for 2-surfaces in the Minkowski spacetime. It is called the rigidity property of quasi-local mass. The rigidity statement of the positive mass theorem says the vanishing of the ADM mass characterizes a Minkowskian initial data set. 

(3) Right asymptotics. The large sphere limits should recover ADM and Bondi mass in spatial and null infinity, respectively. There are 
also expected properties for small sphere limits. 

(4) Good quantitative control. Either along spacelike or null direction (for example, 
for applications to the Penrose inequality), or along timelike direction (for applications to the Einstein evolution equation).  However, it is not expected that a straightforward additivity property should hold because the gravitational binding energy could be negative.

 There are basically four different approaches to the quasi-local mass problem:

(1) Quasi-localization of ADM mass. This was initiated by Bartnik \cite{Bartnik2} by minimizing ADM mass among various asymptotically flat extensions of the finite region. 

(2) Twistor or spinor method by Penrose \cite{Penrose2}, Dougan-Mason \cite{Dougan-Mason}, Ludvigsen-Vickers \cite{LV} et al.

(3) Hamilton-Jacobi method by Brown-York \cite{by1, by2}, Hawking-Horowitz \cite{hh}, Kijowski \cite{ki}, Epp \cite{Epp}, Booth-Mann \cite{Booth-Mann} and Liu-Yau \cite{Liu-Yau}.

(4) The Hawking mass by Hawking \cite{Hawking}, Geroch, Jang, and Wald.

\subsection{The Hawking mass and the Brown-York-Liu-Yau mass}

 In the interest of time, I will focus on the last two types of quasi-local mass and refer other types to the survey paper \cite{Szabados}. These two definitions are more closely related to each other and can be treated in a uniform manner. In fact, both can be defined in terms of the mean curvature vector field.  Given a spacelike 2-surface $\Sigma$ in a spacetime, there exists a unique normal vector field $\vec{H}$ such that \[\delta_V|\Sigma|=-\int_\Sigma \langle \vec{H}, V\rangle\] for any variation vector field $V$ along $\Sigma$, where $|\Sigma|$ is the area of $\Sigma$. 

 When $V$ is a null normal vector field, the integrand $-\langle \vec{H}, V\rangle$ is the null expansion in the direction of $V$.  That $\vec{H}$ is closely related to gravitational energy can be seen from the Penrose singularity theorem which states that spacetime singularity formation can be predicted by the existence of a trapped surface on which both null expansions are negative.

 In the following, we shall assume that $\vec{H}$ is spacelike and $|\vec{H}|>0$.  If $\Sigma$ lies in an initial data set $(M, g, p)$ then $|\vec{H}|^2=k^2-(tr_\Sigma p)^2$, where $k$ is the mean curvature of $\Sigma$ in $M$. 
In the time-symmetric case, i.e. an initial data set with $p=0$, we replace $|\vec{H}|$ by $k$ in the following definitions
\eqref{m} and \eqref{M}.

 The Hawking mass is defined to be: 
\begin{equation}\label{m}m=\sqrt{\frac{|\Sigma|}{16\pi}}(1-\frac{1}{16\pi}\int_\Sigma |\vec{H}|^2),\end{equation} where $|\Sigma|$ is the area of $\Sigma$. 

The Hawking mass has the amazing monotone property (Eardley, Geroch, Jang-Wald \cite{Jang-Wald}) along the inverse mean curvature flow, which is instrumental in Huisken-Ilmanen's \cite{HI} proof of the Riemannian Penrose inequality for a single black hole (see Bray \cite{Bray2001} for the multiple black holes case). The Hawking mass also satisfies 
 monotonicity on asymptotically null or hyperbolic hypersurfaces \cite{Sauter} or in spacetime \cite{Fr, BHMS, BJ}. However, it seems that the limit of the Hawking
 mass approaches the total mass only along a surface foliation that expands to a 2-sphere of constant curvature after rescaling \cite{Neves}.

The Hawking mass is shown to be positive on stable CMC surfaces on time-symmetric slices  by Christdoulou-Yau \cite{Christodoulou-Yau}.

 Another important construction of quasilocal mass was based on the Hamilton-Jacobi method.
The Brown-York-Liu-Yau (BYLY) mass is defined to be:
\begin{equation}\label{M} M=\frac{1}{8\pi}(\int_\Sigma H_0-\int_\Sigma |\vec{H}|)\end{equation} where $H_0$ is the mean curvature of the isometric embedding of the induced metric of $\Sigma$ into $\R^3$, which exists and is unique if  the Gauss curvature is positive by the theorem of Nirenberg \cite{n} and Pogorelov \cite{Pogorelov}. 

 In the time-symmetric case, when $|\vec{H}|$ is replaced by the mean curvature $k$ of $\Sigma$ with respect to an enclosed spacelike region, the Brown-York mass is proven to be positive by Shi-Tam \cite{Shi-Tam} under the condition that the scalar curvature of the enclosed regions is non-negative. The Brown-York mass is gauge dependent. Liu-Yau \cite{Liu-Yau} defined the above mass \eqref{M} that is gauge independent using $|\vec{H}|$ and prove the positivity under the dominant energy condition. 
 
\subsection{Quasi-local mass on spheres of symmetry on a spherically symmetric spacetime}
The metric on a spherically symmetric spacetime takes the form:
\[g_{ab} dx^a dx^b+r^2 d\Omega^2, a, b=0,1 \] where $d\Omega^2=d\theta^2+\sin^2\theta d\phi^2$ is the standard metric on $S^2$ in polar coordinates $(\theta, \phi)$. $SO(3)$ acts on the spacetime by isometry. 
The radius $r$ of an $SO(3)$ orbit, becomes a function on the quotient manifold $Q$ with the Lorentz $(1, 1)$ metric $g_{ab} dx^a dx^b$. Each point $p\in Q$ represents an orbit, a metric round two sphere $\Sigma (p)$ with radius $r(p)$. The mean curvature vector of a sphere $\Sigma (p)$ is

\[\vec{H}=-\frac{2}{r} g^{ab} \partial_a r \frac{\partial}{\partial x^b}=-\frac{2}{r} \nabla r ,\] evaluating at $p$.

Suppose the mean curvature vector of $\Sigma(p)$  is spacelike, the Hawking mass takes the form  
\[m(p)=\frac{r}{2}(1-|\nabla r|^2)\] while the BYLY mass takes the form
\[M(p)=r(1-|\nabla r|).\]

At the horizon of radius $r_0$, where $\nabla r=0$, we have $m(r_0)=\frac{r_0}{2}$. On the Schwarzchild spacetime, $m$ is a constant $m(\infty)=m(r_0)$. On any spherically symmetric spacetime, the relation between $m$ and $M$ is 
\[m(p)=M(p)-\frac{M^2(p)}{2r(p)}.\] In general, $m\leq M$, but they have the same limit as long as $\frac{M^2}{r}\rightarrow 0$ as $r\rightarrow \infty$. 

\subsection{The rigidity property}

 We exam the values of $m$ and $M$ on surfaces in $\R^{3,1}$. 
 A straightforward calculation shows that for a surface in a light cone of $\R^{3,1}$ that encloses the vertex, the Gauss curvature $K=\frac{1}{4}|\vec{H}|^2$ \cite{ost}. On the other hands, for a surface in $\R^3$, the Gauss equation shows $K=\frac{1}{4}|\vec{H}|^2-\frac{1}{4}(\lambda_1-\lambda_2)^2$ where $\lambda_1$ and $\lambda_2$ are the two principal curvatures. 
 
 Therefore, if $\Sigma$ is in a light cone in $\R^{3,1}$ with positive Gauss curvature, then $m=0$ but 
 \[8\pi M=\int_\Sigma (\sqrt{\lambda_1}-\sqrt{\lambda_2})^2>0\] unless $\Sigma$ is a metric round sphere.

For a surface $\Sigma\subset \R^3$, a similar manipulation shows that $m(\Sigma)<0$ unless $\Sigma$ is a metric round sphere.  We shall see in the next section that $M(\Sigma)=0$ for a surface $\Sigma$ in $\R^3$ with positive Gauss curvature by the uniqueness of isometric embedding. Therefore neither of them satisfy the rigidity property of quasi-local mass. 

\section{The new definition of Wang-Yau}

One explanation why the 
BYLY mass does not satisfy the rigidity property is that: for a surface in $\R^{3,1}$, we should look for isometric embeddings into $\R^{3,1}$, instead of into $\R^3$. However, there are good reasons why such isometric embeddings have not been considered and we shall explain in the following.

\subsection{Isometric embeddings into $\R^3$ and $\R^{3,1}$}

Consider a Riemannian metric $\sigma$ on $S^2$, $X:S^2\rightarrow \R^3$ is an isometric embedding of $\sigma$ if $\langle dX, dX\rangle=\sigma$, or in local coordinates $\sum_{i=1}^3 \partial_a X^i \partial_b X^i=\sigma_{ab}, a, b=1, 2$ if $X=(X^1, X^2, X^3)$.  The fundamental theorem is the following solution of the Weyl problem by Nirenberg \cite{n} and Pogorelov \cite{Pogorelov}, independently.

\begin{theorem}
Any Riemannian metric of positive Gauss curvature on $S^2$ can be uniquely isometrically embedded into $\R^3$ up to rigid motions. 
\end{theorem}

 The uniqueness part was proved  earlier by Cohn-Vossen \cite{CV}. The following sketch of proof is closely related to the form of the quasi-local mass by 
 the Hamilton-Jacobi method. Let $X_1$, $X_2$ be two isometric embeddings of $\sigma$ into $\R^3$. Let $\nu_i, H_i, h_i$ be
the respective outward unit normals, mean curvatures, and second fundamental forms of $X_i, i=1, 2$. Then one derives
\[\int_{\Sigma_1} H_1-\int_{\Sigma_2} H_2= 2\int_{\Sigma_1} \det (h_1-h_2) \langle X_1, \nu_1\rangle\]
This is referred to as Herglotz integral formula and the proof generalizes the Minkowski formula $\int_\Sigma H_1=2\int_\Sigma K\langle X_1, \nu_1\rangle$.  One considers the integral \[\int_\Sigma \nabla_a [(h^{ab}-H\sigma^{ab})\langle X, \partial_b\rangle]=0\] and use 
$\nabla_a (h^{ab}-H\sigma^{ab})=0$ (essentially the Codazzi equation or the divergence free property of the conjugate momentum) and 
$\nabla_a\langle X, \partial_b\rangle=\sigma_{ab}+h_{ab}\langle X, \nu\rangle$. The uniqueness theorem then follows from the following result whose proof is left as an exercise: If two $2\times 2$ symmetric positive definite matrixes $h_1, h_2$ have the same determinants then $\det (h_1-h_2)\geq 0$ and the equality
holds if and only if $h_1=h_2$. One approach is to identity the 3-dimensional Minkowski space $\R^{2,1}$ with the space of $2\times 2$ symmetric matrices with the determinant function as the Lorentz metric. The assumption implies that $h_1$ and $h_2$ lie on the same branch of an hyperbola as a level set of the determinant function. 

Although the isometric embedding problem for compact Riemannian manifolds is completely resolved by 
Nash, the issue of uniqueness remains largely untouched except for this case.

 For the isometric embedding problem into $\R^{3,1}$, notice that there are four unknowns (coordinate functions) and three equations. A priori, there should
be at least one functional degree of freedom. This can be identified as the time function with respect to the surface as follows.

 Given any Riemannian metric $\sigma$ and any function $\tau$ on $S^2$, as long as the Gauss curvature of $\sigma+d\tau\otimes d\tau$ is positive, we can embed it into $\R^3$ as a convex surface $\hat{\Sigma}$ by the theorem of Nirenberg and Pogorelov. The graph of $\tau$ over $\hat{\Sigma}$ defines an embedding $X:\Sigma\rightarrow \R^{3,1}$  whose induced metric is $\sigma$. 

 We can replace $\R^3$ by the orthogonal complement of any unit timelike translating Killing field $T_0\in \R^{3,1}$, the time function with respect to $T_0$ is then given by $\tau=-\langle X, T_0\rangle_{\R^{3,1}}$. We shall consider such a pair $(X, T_0)$ and  $T_0$ will play the role of an observer. 

Here is another exercise: Given any closed embedded surface $\Sigma$ in $\R^{3,1}$, the projection of $\Sigma$ onto the orthogonal complement of any $T_0$ is a
closed immersed surface. If furthermore $\Sigma$ bounds a spacelike hypersurface in $\R^{3,1}$, then the projection is embedded. 

\subsection{A sketch of the program}
Since there is one functional degree of freedom, we raise the following question:
 \begin{question}
 Given a physical surface $\Sigma$ in a physical spacetime, among all images of isometric embeddings of the induced metric into  $\R^{3,1}$,
is there a ``best match" of the physical one? 
\end{question}

 To formulate our question more clearly, we take the induced metric $\sigma$ and the mean curvature vector fields ${H}$ of the surface, which is assumed to be spacelike. 
We can reflect ${H}$ along the outgoing light cone in the normal bundle (assume $\Sigma$ bounds a spacelike
region, so there is a distinguished outgoing direction) and obtain a timelike normal vector field $J$. The connection one-form in mean curvature gauge $\alpha_H$ is defined to be 
\[\alpha_H(\cdot)=\langle \nabla^N_{(\cdot)} \frac{J}{|H|}, \frac{H}{|H|}\rangle.\]  This is a new ingredient considered in the Wang-Yau definition in which not just the norm but the direction of the mean curvature vector field ${H}$ is taken into account. 

\begin{question}
 Given the physical data $(\sigma, |H|, \alpha_H)$ is there a pair $(X, T_0)$, where $X$ is an isometric embedding of $\sigma$ into $\R^{3,1}$ and $T_0$ is a unit future timelike Killing fields in $\R^{3,1}$,  that gives  a ``best match" for $(\sigma, |H|, \alpha_H)$?
\end{question}
The hope is that if the data is from a surface in $\R^{3,1}$, then the best matches are just rigid motions of the original 
configuration in $\R^{3,1}$.  However, 
if the data is from a surface of a physically reasonable spacetime, there should be a somewhat unique best match $(X, T_0)$ which allows us to read off  the quasi-local mass and energy-momentum (and other conserved quantities) from the difference
of the geometries.

 We adopt a variational approach and here is a sketch of the program  

1. Given a pair  $(X, T_0)$, define a quasi-local energy $E(X, T_0)$ from the Hamilton-Jacobi analysis and a canonical gauge.

2. Prove the positivity of $E(X, T_0)$ under  convexity conditions on $(X, T_0)$. 

3. Derive the Euler-Lagrange equation of $E(X, T_0)$. 

4. $E(X, T_0)$ can be defined in terms of a function $\rho$ and a one-form $j_a$ on $\Sigma$. 
\[E(X, T_0)=\frac{1}{8\pi}\int_\Sigma (\rho+j_a \nabla_a \tau),\] see \cite[Section 2]{Chen-Wang-Yau3} for the definitions of $\rho$ and $j_a$. $\rho$ is also defined in \eqref{rho} in Section 6. 

The Euler-Lagrange equation becomes $\nabla^a j_a=0$ and the quasi-local energy of a critical point $(X, T_0)$ is simply
$\frac{1}{8\pi}\int_\Sigma \rho$, $\rho$ is called the quasi-local mass density of the critical point. 
We then prove the following stability theorem. 

 \begin{theorem} \cite{Chen-Wang-Yau2} If a critical point $(\bar{X}, \bar{T}_0)$ of $E(X, T_0)$ has positive density $\rho$ and the projection of the image of $\bar{X}$ onto the orthogonal complement of $\bar{T}_0$
is a convex surface, then $(\bar{X}, \bar{T}_0)$ is  energy-minimizing (stable).
\end{theorem}

The last statement is considered as a local uniqueness theorem of the optimal isometric embedding system:
\[\begin{cases} 
\langle dX, dX\rangle_{\R^{3,1}}&=\sigma\\
div_\sigma j=0
\end{cases}\] Note that this is now a system of four unkowns and four equations 

\subsection{Hamilton-Jacobi analysis of the Einstein-Hilbert action}
 For a spacetime domain $D^4$, consider the Einstein-Hilbert action \[\frac{1}{16\pi} \int_D \,^{(4)}R+\frac{1}{8\pi}\int_{\partial D} K+\int_D L(g, \Phi).\] 
Brown-York \cite{by1, by2} and Hawking-Horowitz \cite{hh} derived a surface Hamiltonian on a 2-surface $\Sigma$ in spacetime. The expression
depends on  $t^\nu$, a timelike unit vector fields (an observer, not necessarily normal to the surface) along the surface $\Sigma$, and $u^\mu$, a timelike
unit normal vector field along $\Sigma$. $u^\mu$ is supposed to be the normal to a spacelike hypersurface $\Omega$ bounded by $\Sigma$. The surface Hamiltonian takes the form:
\[\mathfrak{H}(t^\nu, u^\nu)=-\frac{1}{8\pi} \int_\Sigma [Nk-S^\mu (p_{\mu\nu}-p^\lambda_\lambda g_{\mu\nu}) v^\nu],\]
where $t^\nu=N u^\nu+S^\nu$, $N$ is the lapse and $S^\nu$ is the shift vector. $v^\nu$ is the outward unit normal of $\Sigma$ with respect to $\Omega$, $k$ is the mean curvature of $\Sigma$ with respect to $v^\nu$ ($k=-\langle {H}, v\rangle $, where $H$ is the mean curvature vector of $\Sigma$ in spacetime). $p_{\mu\nu}$ is the 2nd fundamental form of $\Omega$ with respect to $u^\nu$.  The expression indeed only depends on the vector fields $t^\nu$ and $u^\nu$ along $\Sigma$ and not on the enclosed region $\Omega$. 

 To obtain a quasi-local energy, consider an isometric embedding into a reference space and take the difference of the physical Hamiltonian and the reference Hamiltonian, or
\[\mathfrak{H}(t_\mu, u^\mu)-\mathfrak{H}(t_0^\mu, u_0^\mu), \] where $\mathfrak{H}(t_\mu, u^\mu)$ is the physical
one and $\mathfrak{H}(t_0^\mu, u_0^\mu)$ is the reference one.

The questions now are the choice of the reference space and the choices of $(t^\mu, u^\mu)$ and $(t_0^\mu, u_0^\mu)$. Both Brown-York and Liu-Yau \cite{Liu-Yau} consider isometric embeddings into the reference space $\R^3$ using the theorem of 
Pogorelov and Nirenberg. Both take $t_0^\mu=u_0^\mu=\frac{\partial}{\partial t}$ and $t^\mu=u^\mu$. The Brown-York definition depends on a spacelike region the physical surface bounds, and $u^\mu$ is the timelike unit normal of the spacelike region. Liu-Yau takes $u^\mu$ such that $H^\mu u_\mu=\langle H, u\rangle=0$, where $H$ is the mean curvature vector of the physical surface. 

\subsection{Wang-Yau quasi-local energy}

 Consider an isometric embedding $X$ into $\R^{3,1}$ and a timelike translating unit Killing field $T_0$ in $\R^{3,1}$. The Wang-Yau prescription is to take $t_0^\nu=T_0$  and $u_0^\nu$ to be the timelike unit normal of $\Sigma$ that is in the direction of $(T_0)^\perp$. $u^\mu$ is chosen such that 
\[H_0^\mu (u_0)_\mu=H^\mu u_\mu.\]  Thus the physical surface in the physical spacetime and the reference surface (the image of $X$) in $\R^{3,1}$ have the same expansion along the respective normals (this is always possible if $H$ is spacelike). 
Finally, $t^\mu=Nu^\mu+S^\mu$ where $N$ and $S$ are the lapse and shift of $t_0^\mu$ with respect to $u_0^\mu$. 
\begin{center}
\includegraphics[width=4in]{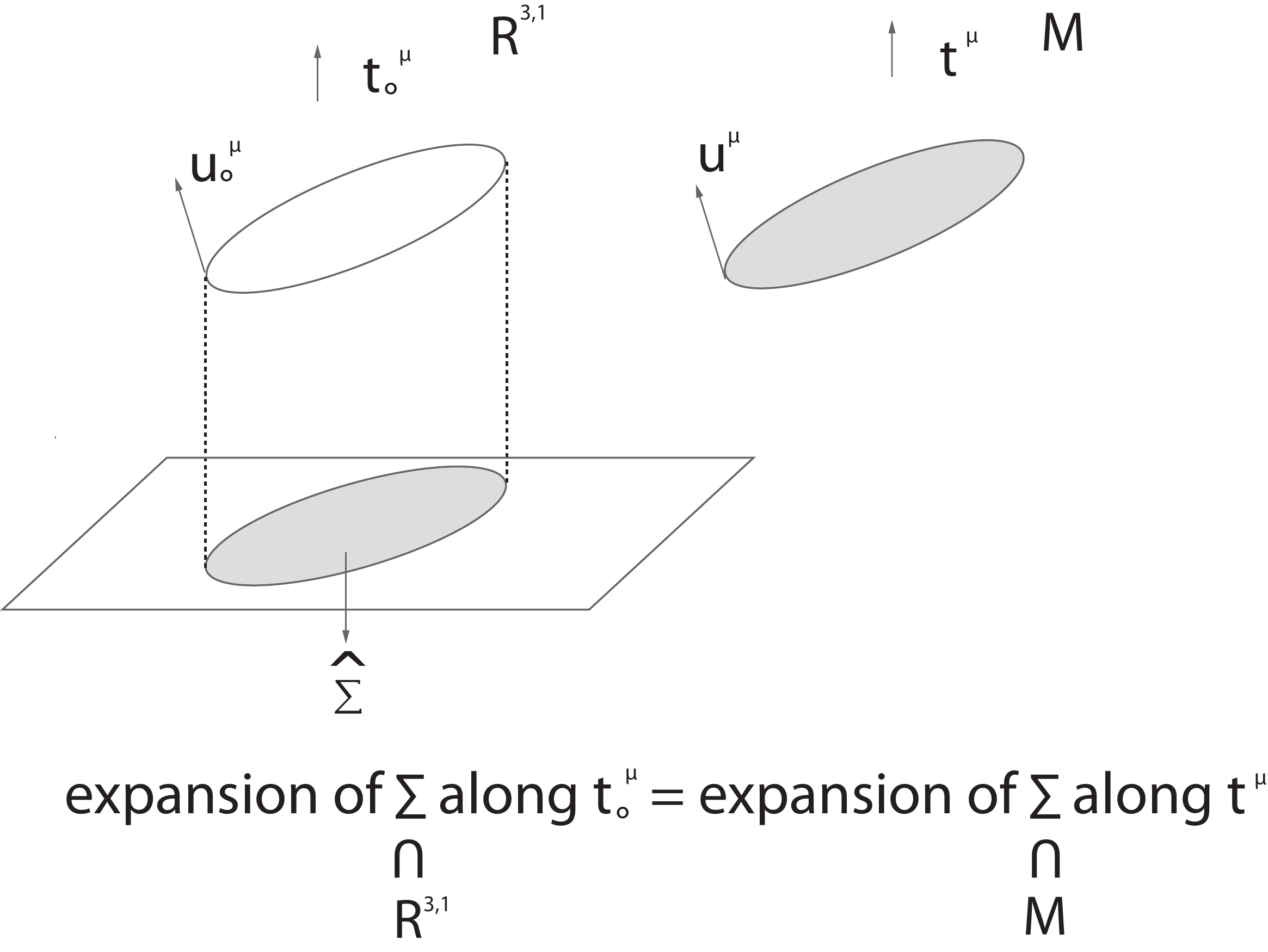}
\end{center}
 \subsection{A gravitational conservation law}

Suppose $u^\mu$ is in the unit timelike normal in the direction of $(t^\mu)^\perp$ in spacetime, and suppose both are tangent to a timelike hypersurface $B$, then the surface Hamiltonian can be rewritten as
\[\mathfrak{H}(t^\mu, u^\mu)=\int_\Sigma \pi_{\mu\nu} t^\mu u^\nu,\] where $\pi_{\mu\nu}$ is the conjugate momentum of $B$. 
Suppose $N$ is a vacuum spacetime ($\nabla^\mu \pi_{\mu\nu}=0$) and $t^\mu$ is Killing then 

\[\int_{\partial B} \pi_{\mu\nu} t^\mu u^\nu=\int_B \nabla^\mu(\pi_{\mu\nu} t^\nu)=0\]
 We apply this to the reference Hamiltonian term, take the timelike hypsurface in $\R^{3,1}$ that is spanned by $X(\Sigma)$ and $T_0$, and let $B$ be the portion of the timelike hypersurface that is bounded by $X(\Sigma)$ and $\hat{\Sigma}$. 
The conservation law implies that the reference surface Hamiltonian is 
\[\mathfrak{H}(t^0, u^0)=-\frac{1}{8\pi}\int_{\hat{\Sigma}} \hat{H}.\]

To compute the physical Hamiltonian, we derive $T_0=\sqrt{1+|\nabla\tau|^2} u_0-\nabla \tau$, $\langle H_0, T_0\rangle=-\Delta\tau$ and thus $\langle H_0, u_0 \rangle =\frac{-\Delta\tau}{\sqrt{1+|\nabla\tau|^2}}$. 

 Therefore, we obtain another form of the quasi-local energy:
\begin{align}\label{form2} E(X, T_0)=\frac{1}{8\pi} \int_{\hat{\Sigma}} \hat{H}-\frac{1}{8\pi} \int_\Sigma [-\sqrt{1+|\nabla\tau|^2} \langle H, \bar{e}_3\rangle-\langle \nabla^N_{\nabla\tau} \bar{e}_3, \bar{e}_4\rangle\end{align} where the gauge $\bar{e}_3, \bar{e}_4$ along the physical surface is determined by $\langle H, \bar{e}_4\rangle=\frac{-\Delta\tau}{\sqrt{1+|\nabla\tau|^2}}$.
Note that we replace $v^\nu$ and $u^\nu$ by $\bar{e}_3$ and $\bar{e}_4$. 

 Remark:
1. This expression vanishes for a surface in $\R^{3,1}$ by the conservation law. Note that there is a canonical identification between $\hat{\Sigma}$ and $X(\Sigma)$ through the graphical relation. The equality is indeed pointwise in the sense that
\begin{equation}\label{pointwise} \hat{H}=-\langle H, \bar{e}_3\rangle-\frac{1}{\sqrt{1+|\nabla\tau|^2}}\langle \nabla^{\R^{3,1}}_{\nabla\tau} \bar{e}_3, \bar{e}_4\rangle.\end{equation}
The relation between the area forms at the corresponding points is $dA_{\hat{\Sigma}}=\sqrt{1+|\nabla\tau|^2} dA_\Sigma$. 

2. The gauge choice $\bar{e}_3, \bar{e}_4$ with $\langle H, \bar{e}_4\rangle=\frac{-\Delta\tau}{\sqrt{1+|\nabla\tau|^2}}$ maximizes the surface Hamiltonian in the following sense:
\begin{equation}\label{maximizing} \int_\Sigma [-\sqrt{1+|\nabla\tau|^2} \langle H, \bar{e}_3\rangle-\langle \nabla^N_{\nabla\tau} \bar{e}_3, \bar{e}_4\rangle]
\geq \int_\Sigma [-\sqrt{1+|\nabla\tau|^2} \langle H,{e}'_3\rangle-\langle \nabla^N_{\nabla\tau} {e}'_3, {e}'_4\rangle]\end{equation}
for any oriented orthonormal frame $e'_3, e'_4$ of the normal bundle of $\Sigma$ in $N$, see \cite[Section 2]{Wang-Yau2}. 

\section{Positivity of the quasi-local energy}

\eqref{form2} is the form of $E(X, T_0)$ to be used to prove the positivity theorem. Given a physical surface $\Sigma$, $\Sigma=\partial M$ for an initial data set $(M, g, p)$ that satisfies the dominant energy condition, and a pair $(X, T_0)$ in which $X$ is an isometric embedding into $\R^{3,1}$ and $T_0$ is a future timelike translating Killing field in $\R^{3,1}$. We assume that the projection of the image $X(\Sigma)$ onto the orthogonal complement of $T_0$ is a convex surface $\hat{\Sigma}$.  The proof is divided into several steps.

\subsection{Bartnik-Shi-Tam construction}
 Let $\hat{\Sigma}=\partial \hat{\Omega}$, for $\hat{\Omega}\subset \R^3$ a compact convex domain, for which given function $\Xi$ defined on $\hat{\Sigma}$ can we prove $\int_{\hat{\Sigma}} (\hat{H}-\Xi)\geq 0$?

We aim to take \[\Xi= -\langle H, \bar{e}_3\rangle- \frac{1}{\sqrt{1+|\nabla\tau|^2}}\langle \nabla^N_{\nabla\tau} \bar{e}_3, \bar{e}_4\rangle.\] 
Such a theorem can be proved by a Bartnik-Shi-Tam construction. This involves the quasi-spherical construction which was introduced in Bartnik's definition of quasi-local mass \cite{Bartnik2}. 

 Write the flat metric on $\R^3\backslash \hat{\Omega}$ as $dr^2+g_{ab} dx^a dx^b$ where $r$ is the distance function to $\hat{\Sigma}$ and $g_{ab} dx^adx^b$ is the induced metric on $\Sigma_r$, the level set of $r$.  For a function $u$ defined on 
$\R^3\backslash \hat{\Omega}$, consider the new metric $g_u=u^2 dr^2+g_{ab} dx^a dx^b$. The scalar flat equation $R(g_u)=0$ with prescribed $u>0$ on $\Sigma_0=\hat{\Sigma}$ is a quasilinear parabolic equation with a unique smooth solution. 
The new metric $g_u$ on $\R^3\backslash \hat{\Omega}$ is asymptotically flat and 

\[\int_{\hat{\Sigma}} (\hat{H}-\frac{\hat{H}}{u})\geq \text{ ADM energy of } g_u\]

 The proof relies on the following two properties of $e(r)=\int_{\Sigma_r} (\hat{H}_r-\frac{\hat{H}_r}{u})$ where $\hat{H}_r$ is the mean curvature of $\Sigma_r$ with respect to the flat metric:
\[e'(r)\leq 0\] and \[\lim_{r\rightarrow \infty} e(r)=\text{ ADM energy of } g_u.\]

Shi-Tam proved the following theorem which asserts the positivity of the Brown-York mass:
\begin{theorem} \cite{Shi-Tam}
Suppose $\Omega$ is a compact Riemannian 3-manifold with non-negative scalar curvature.
Suppose $\Sigma=\partial \Omega$  has positive Gauss curvature and positive mean curvature $k$ with respect to outward normal of $\Omega$. If $\Sigma$ is isometric to $\hat{\Sigma}\subset \R^3$, then 
\[\int_{\hat{\Sigma}} \hat{H}-\int_\Sigma k\geq 0\] and equality holds if and only if $\Omega$ is flat. 
\end{theorem}

In the proof, the initial value of $u$ on $\hat{\Sigma}=\Sigma_0$ is chosen to be $\frac{\hat{H}}{k}$ in the Bartnik-Shi-Tam construction. It suffices to prove that the ADM energy
of the metric $g_u$ is positive. The idea is to glue the Riemannian manifolds $\Omega$ and $(\R^3\backslash \hat{\Omega}, g_u)$ together along their respective boundaries $\Sigma$ and $\hat{\Sigma}$ (which are isometric) and apply Witten's spinor proof of the positive mass theorem. The proof needs to be adapted to a Lipschitz manifold with distributional non-negative scalar curvature, see \cite{Shi-Tam}.

In the proof of the positivity of the quasi-local energy \eqref{form2}, the following extension is needed (see \cite[Theorem 5.1]{Wang-Yau2}):

\begin{theorem}\label{positivity_thm1}
Suppose $\tilde{\Omega}$ is a compact Riemannian 3-manifold and $\tilde{\Sigma}=\partial \tilde{\Omega}$ has positive 
Gauss curvature. Suppose there exists a vector field $V$ on $\tilde{\Omega}$ such that 
\[R\geq 2|V|^2-2 div V \text{ on } \tilde{\Omega}\] and 
\[k> \langle V, \tilde{\nu}\rangle \text{  on } \tilde{\Sigma}\] where $k$ is the mean curvature of $\tilde{\Sigma}$ with
respect to the outward normal $\tilde{\nu}$. If $\tilde{\Sigma}$ is isometric to $\hat{\Sigma}\subset \R^3$, then
\[\int_{\hat{\Sigma}}\hat{H}\geq \int_{\tilde{\Sigma}} (k-\langle V, \tilde{\nu}\rangle).\]
\end{theorem}

The remaining question is to construct such data and to relate the term $\int_\Sigma (k-\langle V, \tilde{\nu}\rangle)$ to the physical Hamiltonian in \eqref{form2}.

\subsection{Jang-Schoen-Yau equation}
Here is an important observation of Schoen-Yau in their proof of the positive mass theorem \cite{SY2}. A natural question arises in their consideration as how one can recognize an initial data set in the Minkowski space.
 A spacelike hypersurface in $\R^{3,1}$ is a graph defined by $t=f(x^1, x^2, x^3)$ such that the $|\nabla f|<1$ where $f$ is a function defined on a domain in $\R^3$ and $|\nabla f|$ is with respect to the flat metric on $\R^3$. We can identify the domain on which $f$ is defined with the graph over it and consider $f$ as a function defined on the graph, and the second fundamental form of the graph can be expressed in term of the Hessian $f$.  
 
 Therefore for $(M, g, p)$ is a Minkowskian data, there exists a function $f$ (the defining function of $M$) on $M$ such that 
\[p_{ij}=\frac{D_i D_j f}{\sqrt{1+|Df|^2}}\] and that $g_{ij}+f_i f_j$ is a flat metric where $D_i D_j f$ and $|Df|^2$ are with respect to $g$. This is simply the formula of the second fundamental form of a graph, except the metric on the graph is used to express the derivatives. 

 In general, to each $T_0$ there exists an $f$ that corresponds to the projection along $T_0$. However,
if $M$ is compact with boundary, there is a unique one subject to each boundary condition. 

For a general initial data $(M, g, p)$, consider the following Jang's equation for $f$. 
\begin{equation}\label{Jang} (g^{ij}-\frac{f^i f^j}{1+|Df|^2})(\frac{D_iD_j f}{\sqrt{1+|Df|^2}}-p_{ij})=0\end{equation} where $f^i=g^{ik}f_k$. 

 The graph of $f$ in $M\times \R$ is a Riemannian manifold with induced metric $g_{ij}+f_i f_j$. Equation \eqref{Jang} is a quasilinear elliptic equation for $f$. In fact if $(M, g, p)=(\R^3, \delta, 0)$, the equation 
becomes the minimal surface equation
\[\text{div}(\frac{\text{ grad }f}{\sqrt{1+|\text{ grad } f|^2}})=0.\] The equation has geometric interpretations in terms of MOTS.

The following important inequality was derived by Schoen-Yau:
\begin{theorem} \cite{SY2}
Given an initial data set $(M, g, p)$ that satisfies the dominant energy condition. Suppose a solution of the Jang
equation exists, then on the graph of the solution, $\tilde{\Omega}$ in $M\times \R$ there exists a vector field $V$ such that 
\begin{align}\label{RX}R-2|V|^2+2div V\geq 0,\end{align} where $R$ is the scalar curvature on the graph and $|\cdot|$ and $div$ are respect
to the induced metric on the graph. 
\end{theorem}

$V$ can be directly computed from the graph of $f$. 
The condition \eqref{RX} indeed guarantees that the graph is conformally scalar flat, i.e. it can be turned into a
metric of zero scalar curvature after a conformal change.

We take the initial data set $(M, g, p)$ that the physical surface $\Sigma$ bounds, consider a pair $(X, T_0)$ with $\tau=-\langle X, T_0\rangle$.
Solve the Jang equation with the boundary data $f=\tau $ on $\Sigma$.  The graph over the boundary, denoted by $\tilde{\Sigma}$, has the induced metric $\sigma+d\tau\otimes d\tau $, and is isometric to the surface $\hat{\Sigma}$ in $\R^3$.  Let $\tilde{\Omega}$ denotes the graph of the solution of the Jang equation in $M\times \R$ with boundary data $\tau$. 
If $k-\langle V, \tilde{\nu}\rangle>0$ on $\tilde{\Sigma}$ the boundary of the graph, Theorem \ref{positivity_thm1} is applicable. We can construct the metric $g_u$ on $\R^3\backslash \hat{\Omega}$ with the initial value $u=\frac{\hat{H}}{k-\langle V, \tilde{\nu}\rangle}$ on $\Sigma_0=\hat{\Sigma}$.  Since $\tilde{\Sigma}$ and $\hat{\Sigma}$ are isometric, we can glue together 
$\tilde{\Omega}$ with the induced metric and $(\R^3\backslash \hat{\Omega}, g_u)$ along them. 
  But how is this result 
related to what we really want to prove, namely the positivity of \eqref{form2}? It turns out the quantity $k-\langle V, \tilde{\nu}\rangle$ on $\tilde{\Sigma}$ can be related to a surface Hamiltonian density on $\Sigma$ through the natural identification, and we can close the argument by the maximizing property \eqref{maximizing} of the canonical gauge. 

\begin{theorem} Through the natural identification of points on $\tilde{\Sigma}$ and ${\Sigma}$, we have
\[k-\langle V, \tilde{\nu}\rangle=-\langle H, e'_3\rangle-\frac{1}{\sqrt{1+|\nabla \tau|^2}}\langle \nabla^N_{\nabla\tau} 
e'_3, e'_4\rangle\] where 
$e'_3=\cosh\phi e_3+\sinh \phi e_4$ and $\sin\phi=\frac{-e_3(f)}{\sqrt{1+|\nabla\tau|^2}}$. 
\end{theorem}
Here the $e_4$ is the future timelike unit normal  of $(M, g, p)$ in spacetime and $e_3$ is the outward spacelike unit normal of $\Sigma$ as the boundary of $M$.
When $(M, p, g)$ is a Minkowskian initial data set, this recovers the pointwise conservation law \eqref{pointwise}. The left hand side is the mean curvature $\hat{H}$ of the surface  $\hat{\Sigma}$ in $\R^3$ and the right hand side the surface Hamiltonian
on $\Sigma$ with respect to the canonical gauge.

We summarize the positivity theorem of the quasi-local energy:
\begin{theorem}
The quasilocal energy of a physical 2-surface $\Sigma$ with data $(\sigma, |H|, \alpha_H)$ with respect to $(X, T_0)$  is non-negative if the physical surface 
bounds an initial data set $(M, g, p)$ that satisfies the dominant energy condition and 

(1) $\sigma+d\tau\otimes d\tau$ is of positive Gauss curvature where $\tau=-\langle X, T_0\rangle$.

(2) The Jang equation over $M$ with boundary data $\tau$ admits a smooth solution. 

(3) The boundary term $k-\langle X, \tilde{\nu}\rangle$ on the boundary of the graph of the solution is positive. 

If the energy is zero, then $(M, g, p)$, and thus $\Sigma$, is a Minkowskian data.

\end{theorem}

 We remark that the Jang equation is solvable if $\Sigma=\partial M$ has spacelike mean curvature and there is no MOTS or MITS
in $(M, g, p)$, see \cite[Theorem 5]{Chen-Wang-Yau2}.

Suppose the physical surface $\Sigma$ is in the Minkowski spacetime, which is given by the isometric embedding $\bar{X}$. By the definition of the quasi-local energy and the conservation law,  $E(\bar{X}, \bar{T}_0)=0$ for any $\bar{T}_0$. The positivity theorem implies that $(\bar{X}, \bar{T}_0)$ is a stable critical point for the quasi-local energy as long as the projection of $\Sigma$ along the direction of $T_0$ is a convex surface.

\begin{corollary}
Suppose $\Sigma\subset \R^{3,1}$ is given by an embedding $\bar{X}$ with induced metric $\sigma$. Suppose for a unit timelike translating Killing field $\bar{T}_0$, $\sigma+d\bar{\tau}\otimes d\bar{\tau}$ is of positive Gauss curvature for $\bar{\tau}=-\langle \bar{X}, \bar{T}_0\rangle$, then $(\bar{X}, \bar{T_0})$ is a stable critical point for $E$, i.e. for any $(X, T_0)$ close enough to $(\bar{X}, \bar{T}_0)$, one has
\[E(X, T_0)\geq E(\bar{X}, \bar{T}_0)\] and the equality holds if and only if $(X, T_0)$ is related to $(\bar{X}, \bar{T}_0)$ through a Lorentz transformation.
\end{corollary}

\section{Optimal isometric embedding and the stability theorem}

\subsection{Optimal isometric embedding system}
 Which $(X, T_0)$ is the best match of the physical 2-surface? For example, take a surface data $(\sigma, |H|, \alpha_H)$ that is the data on a surface in the Minkowski space, how does it find 
itself as the best match? Recall that in relativity, the energy depends on observers and the rest mass is the minimum of energy seen by all observers. The idea is thus to minimize $E(X, T_0)$. In order to compute the variation and derive the Euler-Lagrangian equation, we rewrite the physical surface Hamiltonian in terms of the physical data $|H|$ and $\alpha_H$. The quasi-local energy with respect to $(X, T_0)$ is 
\[\begin{split}&E(X, T_0)=\frac{1}{8\pi}\int_{\widehat{\Sigma}} \widehat{H} -\frac{1}{8\pi}\int_\Sigma \left[\sqrt{1+|\nabla\tau|^2}\cosh\theta|{H}|-\nabla\tau\cdot \nabla \theta -\alpha_H ( \nabla \tau) \right]\end{split}\] where $\nabla$ and $\Delta$ are the gradient and Laplace operator of $\sigma$ respectively, $\tau=-\langle X, T_0\rangle$ is considered as a function on the 2-surface $\Sigma$,
$|\nabla \tau|^2=\sigma^{ab}\nabla_a \tau\nabla_b\tau$, $\Delta \tau=\nabla^a\nabla_a \tau$, and \begin{equation}\label{theta}\theta=\sinh^{-1}(\frac{-\Delta\tau}{|H|\sqrt{1+|\nabla\tau|^2}}).\end{equation}

 The variation of the total mean curvature with respect to a metric deformation is (see \cite[Section 6]{Wang-Yau2}) \[\delta_{\delta\hat{\sigma}} \int_{\hat{\Sigma}} \hat{H} =-\frac{1}{2}\int_{\hat{\Sigma}} (\hat{h}_{ab}-\hat{H}\hat{\sigma}_{ab})\delta{ \hat{\sigma}^{ab}}.\]

It is clear that the variation has to be related to the conjugate momentum because a reparametrization leaves the total mean curvature invariant and the conjugate momentum is the only natural divergence free $(0, 2)$ tensor that satisfies this property. A variation of $\tau$ induces a variation of the metric $\hat{\sigma}=\sigma+d\tau\otimes d\tau$ on $\hat{\Sigma}$. The variation of the right side can be computed
by holding the physical term $|H|$ and $\alpha_H$ and the final result for the Euler-Lagrange equation is:
\[   -(\widehat{H}\hat{\sigma}^{ab} -\hat{\sigma}^{ac} \hat{\sigma}^{bd} \hat{h}_{cd})\frac{\nabla_b\nabla_a \tau}{\sqrt{1+|\nabla\tau|^2}}+ div_\sigma (\frac{\nabla\tau}{\sqrt{1+|\nabla\tau|^2}} \cosh\theta|{H}|-\nabla\theta-\alpha_{H})=0.
\]

 Together with the isometric embedding equation $\langle dX, dX\rangle_{\R^{3,1}}=\sigma$, this forms the so-called optimal isometric
embedding system introduced in \cite{Wang-Yau2}. A solution $(X, T_0)$ of this system is a critical point of $E(X, T_0)$. For example, if $div_\sigma \alpha_H=0$ or the surface is timeflat \cite{BJ, Chen-Wang-Wang}, then $\tau=0$ or isometric embedding into $\R^3$
is a solution of this system.  A sphere of symmetry in a spherically symmetric spacetime, or an axial symmetric surface
in the maximal slice of the Kerr spacetime is timeflat. 

\subsection{Stability of critical points}
We saw in the previous section that for a surface in $\R^{3.1}$ given by an embedding $\bar{X}$, $(\bar{X}, \bar{T}_0)$ is a stable critical point of the quasi-local energy if the projection along $T_0$ is convex. For a general critical point, we have the following stability theorem: 

\begin{theorem} \cite{Chen-Wang-Yau2}
Let $(\sigma, |H|, \alpha_H)$ be the data of a spacelike surface $\Sigma$ in a general spacetime. Suppose that $(\bar{X}, \bar{T}_0)$ is a critical point of the quasi-local energy $E(X, T_0)$ and that the corresponding quasilocal mass density $\rho$ of $(\bar{X}, \bar{T}_0)$  is positive, then $(\bar{X}, \bar{T}_0)$ is a local minimum for  $E(X, T_0)$.
\end{theorem}

The quasi-local mass density $\rho$, see equation (2.2) of \cite{Chen-Wang-Yau3}, of $(\bar{X}, \bar{T}_0)$ is defined to be 

 \begin{equation} \label{rho} \begin{split}\rho &= \frac{\sqrt{|H_0|^2 +\frac{(\Delta \tau)^2}{1+ |\nabla \tau|^2}} - \sqrt{|H|^2 +\frac{(\Delta \tau)^2}{1+ |\nabla \tau|^2}} }{ \sqrt{1+ |\nabla \tau|^2}}, \end{split}\end{equation} where $H_0$ is the mean curvature vector filed of $\bar{X}(\Sigma)$ and $\tau=-\langle \bar{X}, \bar{T}_0\rangle$. Thus the positivity of $\rho$ is equivalent to $|H_0|>|H|$.

 The special case when  the isometric embedding $\bar{X}$ is into $\R^3\subset \R^{3,1}$ was proved by Miao-Tam-Xie \cite{MTX}. They derived explicitly the second variation operator in this special case and showed the operator is indeed positive definite by Reilly's formula. 
 
The general case relies on the following comparison theorem. 

\begin{theorem} \cite{Chen-Wang-Yau2} 
Under the assumption of the above theorem, for any  $(X, T_0)$  with $\tau=-\langle X, T_0\rangle$ such that $ \sigma + d \tau \otimes d \tau$ has positive Gaussian curvature, we have
\begin{equation}\label{comparison} E(\Sigma,X, T_0) \ge E(\Sigma,\bar{X}, \bar{T}_0) +E(\bar{X}(\Sigma), X, T_0).\end{equation}
Moreover, equality holds if and only if $\tau-\tau_0$ is a constant.  
\end{theorem}
The last term is computed by taking the image of $\bar{X}$ in the Minkowski spacetime as a physical surface and evaluating its quasi-local energy with respect to another pair $(X, T_0)$. 

The stability theorem allows us to solve the optimal isometric embedding system in any perturbative configuration. Several applications of quasi-local energy rely on this theorem, see \cite{Chen-Wang-Yau1, Chen-Wang-Yau3, Chen-Wang-Yau4}.


\begin{thebibliography}{99}

\bibitem{Arnowitt-Deser-Misner} R. Arnowitt, S. Deser\ and\ C. W. Misner, {\it The dynamics of general relativity}, in Gravitation: An introduction to current research, 227--265, Wiley, New York.

\bibitem{Bartnik1} R. Bartnik, \textit{The mass of an asymptotically flat manifold,} Comm. Pure Appl. Math. 39 (1986), no. 5, 661--693.
\bibitem{Bartnik2}R. Bartnik, {\it New definition of quasi-local mass}, Phys. Rev. Lett. {\bf 62} (1989), no.~20, 2346--2348.

\bibitem{BVM}  H. Bondi,  M. G. J. van der Burg, and  A. W. K. Metzner, {\em Gravitational waves in general relativity. VII. Waves from axi-symmetric isolated systems},
Proc. Roy. Soc. Ser. A 269 (1962) 21--52.
\bibitem{Booth-Mann} I. S. Booth and R. B. Mann, Phys. Rev. D 59, 064021
(1999).

\bibitem{Bray2001} H. L. Bray, \textit{Proof of the Riemannian Penrose inequality using the positive mass theorem,} J. Differential Geom. 59 (2001), no. 2, 177Ð267.

\bibitem{BHMS} H. Bray, S. Hayward, M. Mars, Marc, and W. Simon, \textit{Generalized inverse mean curvature flows in spacetime,} Comm. Math. Phys. 272 (2007), no. 1, 119Ð138. 

\bibitem{BJ} H. L. Bray and J. L. Jauregui, \textit{Time flat surfaces and the monotonicity of the spacetime Hawking mass,} Comm. Math. Phys. 335 (2015), no. 1, 285Ð307.
\bibitem{by1} J. D. Brown\ and\ J. W. York, Jr., {\it Quasilocal energy in general relativity}, in Mathematical aspects of classical field theory (Seattle, WA, 1991), 129--142, Contemp. Math., 132, Amer. Math. Soc., Providence, RI.
\bibitem{by2} J. D. Brown and J. W.  York,  \textit{Quasi-local energy and conserved charges derived from the gravitational action,} Phys. Rev. D (3) \textbf{47} (1993), no. 4, 1407--1419.
\bibitem{Chen-Wang-Yau1} P.-N. Chen, M.-T. Wang, and S.-T. Yau, \textit {Evaluating quasi-local energy and solving optimal embedding equation at null infinity}, Comm. Math. Phys. \textbf{308} (2011), no.3, 845--863.
\bibitem{Chen-Wang-Yau2}  P.-N. Chen, M.-T. Wang, and S.-T. Yau, \textit{Minimizing properties of critical points of quasi-local energy}, Comm. Math. Phys. \textbf{329} (2014), no.3, 919--935
\bibitem{Chen-Wang-Yau3}  P.-N. Chen, M.-T. Wang, and S.-T. Yau, \textit{Conserved quantities in general relativity: from the quasi-local level to spatial infinity}, Comm. Math. Phys. \textbf{338} (2015), no.1, 31--80.
\bibitem{Chen-Wang-Yau4}  P.-N. Chen, M.-T. Wang, and S.-T. Yau, \textit{Conserved quantities on asymptotically
hyperbolic initial data sets}, arXiv: 1409.1812
\bibitem{Chen-Wang} P.-N. Chen and M.-T. Wang, \textit{Rigidity and minimizing properties of quasi-local mass}, arXiv: 1411.1625

\bibitem{Chen-Wang-Wang} P.-N. Chen, M.-T. Wang, and Y.-K. Wang, \textit{Rigidity of time-flat surfaces in the Minkowski spacetime,} Math. Res. Lett. 21 (2014), no. 6, 1227--1240. 
\bibitem{Christodoulou-Yau}D. Christodoulou\ and\ S.-T. Yau, Some remarks on the quasi-local mass, in {\it Mathematics and general relativity (Santa Cruz, CA, 1986)}, 9--14, Contemp. Math., 71, Amer. Math. Soc., Providence, RI.
\bibitem{Chrusciel86} P. T. Chru\'{s}ciel, 
\textit{A remark on the positive-energy theorem,}
Classical Quantum Gravity 3 (1986), no. 6, L115--L121. 
Corrigendum: Classical Quantum Gravity 4 (1987), no. 4, 1049. 
 The spacetime positive mass theorem in dimensions less than eight
\bibitem{CV} E. Cohn-Vossen, \textit{Zwei S\"{a}tze \"{u}ber die Starrheit der Eifl\"{a}chen}, in Nachrichten von der
Gesellschaft der Wissenschaften zu G\"{o}ttingen, Mathematish-Physikalische Klasse, 1927,
pp. 125--134.
\bibitem{Dougan-Mason} A. J. Dougan and L. J. Mason, \textit{Quasilocal mass constructions with positive energy}, Phys.
Rev. Lett., 67, 2119--2122, (1991).
\bibitem{EHLS} M. Eichmair, L.-H. Huang, D. A. Lee, and R. M. Schoen,  \textit{The spacetime positive mass theorem in dimensions less than eight,} arXiv:1110.2087 
\bibitem{Epp} R. J. Epp, Phys. Rev. D 62, 124018 (2000).
\bibitem{Fr} J. Frauendiener, \textit{On the Penrose inequality,} Phys. Rev. Lett. 87, 101101--1 (2001) 
\bibitem{G} G. W. Gibbons, \textit{Collapsing shells and the isoperimetric inequality for black holes}, Classical Quantum Gravity {\bf 14} (1997), no.~10, 2905--2915.
\bibitem{Hawking} S. W. Hawking, \textit{Gravitational radiation in an expanding universe},
J. Math. Phys. \textbf{9}, 598 (1968).
\bibitem{hh} S. W. Hawking and G. T.  Horowitz,
\textit{The gravitational Hamiltonian, action, entropy and surface terms,}
Classical Quantum Gravity \textbf{13} (1996), no. 6, 1487--1498.
\bibitem{HI} G. Huisken and T. Ilmanen, {\em The inverse mean curvature flow and the Riemannian Penrose inequality.} J. Differential Geom. \textbf{59} (2001), no. 3, 353--437.
\bibitem{Jang-Wald} P. S. Jang and R. M. Wald,
\textit{The positive energy conjecture and the cosmic censor hypothesis, }
J. Mathematical Phys. 18 (1977), no. 1, 41Ð44.
\bibitem{ki} J. Kijowski,  \textit{A simple
derivation of canonical structure and quasi-local
 Hamiltonians in general relativity,} Gen. Relativity Gravitation \textbf{29} (1997), no. 3, 307--343.
\textbf{90} (2003), no. 23, 231102.
\bibitem{Liu-Yau} C.-C. M. Liu and S.-T.  Yau,
\textit{Positivity of quasi-local mass II,} J. Amer. Math. Soc. \textbf{19} (2006), no. 1, 181--204.

\bibitem{LV} M. Ludvigsen and J.A.G. Vickers, J.A.G., \textit{Momentum, angular momentum and their quasi-local
null surface extensions}, J. Phys. A, 16, 1155--1168, (1983).
\bibitem{MTX} P. Miao, L.-F. Tam and N. Xie, {\em Critical points of Wang-Yau quasi-local energy}, Ann. Henri Poincare, {\bf 12} (2011), no. 5, 987--1017. 

\bibitem{ost}N. \'O Murchadha, L. B. Szabados\ and\ K. P. Tod, \textit{Comment on: ``Positivity of quasi-local mass''},  Phys. Rev. Lett. {\bf 92} (2004), no.~25, 259001, 1 p. 
\bibitem{Neves} A. Neves, \textit{Insufficient convergence of inverse mean curvature flow on asymptotically hyperbolic manifolds}, J. Differential Geom. {\bf 84} (2010), no.~1, 191--229.
\bibitem{n} L. Nirenberg, \textit{The Weyl and Minkowski problems in differential geometry in the large,} Comm. Pure Appl. Math. \textbf{6}, (1953), 337--394.

\bibitem{PT} T. Parker and C. H. Taubes, 
\textit{On Witten's proof of the positive energy theorem,}
Comm. Math. Phys. 84 (1982), no. 2, 223Ð-238. 


\bibitem{P1} R. Penrose, {\em Some unsolved problems in classical general relativity}, Seminar on Differential Geometry, pp. 631--668,  Ann. of Math. Stud., 102, Princeton Univ. Press, Princeton, N.J., 1982. 
\bibitem{Penrose2} R. Penrose, \textit{Quasi-local mass and angular momentum in general relativity}, Proc. Roy. Soc. London Ser. A {\bf 381} (1982), no.~1780. 

\bibitem{Pogorelov} A. V. Pogorelov, \textit{Regularity of a convex surface with given Gaussian curvature,} (Russian) Mat. Sbornik N.S. \textbf{31}(73), (1952), 88--103.
\bibitem{Sauter} J. Sauter, \textit{Foliations of null hypersurfaces
and the Penrose Inequality}, Ph. D. thesis (2008), ETH, Zurich, Switzerland 

\bibitem{SY1} R. Schoen and S.-T. Yau, {\em On the proof of the positive mass conjecture in general relativity. } Comm. Math. Phys. {\bf 65} (1979), no. 1, 45--76. 
\bibitem{SY2} R. Schoen and S.-T. Yau, {\em Proof of the positive mass theorem. II}, Comm. Math. Phys. {\bf 79} (1981), no. 2, 231--260. 
\bibitem{Shi-Tam}Y. Shi\ and\ L.-F. Tam, \textit{Positive mass theorem and the boundary behaviors of compact manifolds with nonnegative scalar curvature}, J. Differential Geom. {\bf 62} (2002), no.~1, 79--125.
\bibitem{Szabados} L. B. Szabados, \textit{Quasi-local energy-momentum and angular momentum in general relativity}, Living Rev. Relativity {\bf 12},  (2009).
\bibitem{Tod} K. P. Tod, Penrose's quasi-local mass, in {\it Twistors in mathematics and physics}, 164--188, London Math. Soc. Lecture Note Ser., 156, Cambridge Univ. Press, Cambridge.
\bibitem{Wang-Yau1} M.-T. Wang, and S.-T. Yau, {\em Quasi-local mass in general relativity}, Phys. Rev. Lett. {\bf 102} (2009), no. 2, no. 021101.
\bibitem{Wang-Yau2}  M.-T. Wang, and S.-T. Yau, {\em Isometric embeddings into the Minkowski space and new quasi-local mass}, Comm. Math. Phys. {\bf 288} (2009), no. 3, 919--942. 
\bibitem{Wang-Yau3} M.-T. Wang, and S.-T. Yau, {\em Limit of quasi-local mass at spatial infinity}, Comm. Math. Phys. {\bf 296} (2010), no. 1, 271--283. 
\bibitem{Witten} E. Witten,  {\em A new proof of the positive energy theorem.} 
Comm. Math. Phys. {\bf 80} (1981), no. 3, 381--402. 
\end{thebibliography}
\end{document}